\documentclass{IEEEtran}
\usepackage{cite}
\usepackage{algorithmic}
\usepackage{xcolor}
\usepackage{enumerate}
\usepackage{amsfonts,amsmath,amssymb,amsfonts}
\usepackage{algorithmic}
\usepackage{optidef}
\usepackage{algorithm}
\usepackage{graphicx}
\usepackage{subfigure}
\usepackage[section]{placeins}
\def\BibTeX{{\rm B\kern-.05em{\sc i\kern-.025em b}\kern-.08em
    T\kern-.1667em\lower.7ex\hbox{E}\kern-.125emX}}
\begin{document}
\bibliographystyle{IEEEtran}
\title{Reconfigurable Intelligent Surface Assisted Device-to-Device Communications}

\author{Zelin~Ji, Zhijin~Qin, and  Clive~G.~Parini
\thanks{Part of this work was presented at the IEEE Global Communications Conference 2020 \cite{9322411}.}
\thanks{Zelin~Ji, Zhijin~Qin, and  Clive~G.~Parini are with Queen Mary University of London, London, UK, E1 4NS, email: \{z.ji, z.qin, c.g.parini\}@qmul.ac.uk.}
}
\maketitle

\begin{abstract}
Reconfigurable intelligent surface (RIS) technology is a promising method to enhance wireless communications services and to realize the smart radio environment. In this paper, we investigate the application of RIS in D2D communications, and maximize the sum of the transmission rate of the D2D underlaying networks in a new perspective. Instead of solving similarly formulated resource allocation problems for D2D communications, this paper treats the wireless environment as a variable by adjusting the position and phase shift of the RIS. To solve this non-convex problem, we propose a novel double deep Q-network (DDQN) based structure which is able to achieve the near-optimal performance with lower complexity and enhanced robustness. Simulation results illustrate that the proposed DDQN based structure can achieve a higher uplink rate compared to the benchmarks, meanwhile meeting the quality of service (QoS) requirements at the base station (BS) and D2D receivers.
\end{abstract}

\begin{IEEEkeywords}
Deep reinforcement learning, Device-to-device communication, Non-convex optimization, Reconfigurable intelligent surfaces.
\end{IEEEkeywords}

\section{Introduction}

As one of the key technologies of the fifth-generation (5G) and beyond communication systems, underlaying device-to-device (D2D) communications permit devices to communicate with proximity devices over the licensed spectrum for cellular networks, thus enhancing the communication system performance by reducing the latency, improving energy efficiency (EE), and spectrum efficiency (SE)~\cite{8340813}. D2D communications have been applied in various applications, including the 3rd Generation Partnership Project (3GPP) proximity services~\cite{6807945}, Internet of Things (IoT), vehicle-to-everything (V2X) communications, and wearable communications~\cite{hoyhtya2018review}. According to the expectation of Cisco, the share of D2D links will increase 20 percent from 2018 to 2023~\cite{tensorflow2015-whitepaper}. These various applications pushes the development and implementation of D2D communications. 

There is a rich body of literature that focuses on the resource allocation for D2D communications~\cite{5910123,7579565,7913583}. Recently, a novel approach referred to as the smart radio environment presents a new perspective to enhance the D2D communications. Particularly, the wireless environment can be controllable and programmable. In this way, we can optimize the communication environment and resource allocation for D2D devices simultaneously, thus permitting us to control or eliminate the interference significantly.

One key technique to realize the smart radio environment is reconfigurable intelligent surfaces (RISs)~\cite{di2019smart}, which have attracted extensive attention in wireless communications. Equipped with an array of low-cost passive reflecting elements, the phase shift and reflection amplitude of each RIS element can be adjusted by a controller, enabling it the ability to modify wireless communication environment proactively. Compared with the conventional relays, the advantages of RIS include lowered energy consumption and enhanced system capacity~\cite{8796365}. Although the control signal can be analog using varactors to achieve continuous phase shift~\cite{7510962}, the long response time and low phase accuracy of varactors make it impractical for wireless communications. Theoretical analyses have been provided for multi-bit controlled elements to strike a tradeoff between the system performance and the complexity~\cite{4476079, 8930608}. The performance improvement has been further verified by a RIS-based wireless communication prototype~\cite{9020088}, which motivates us to apply it in D2D communication systems. 

On the other hand, the large number of RIS elements requires optimization approaches with lower complexity. Although typical optimization tools such as exhaustive search can generate the optimal solution, the high computational cost makes it unrealistic for the real-time optimization. Fortunately, machine learning (ML) methods, especially deep learning (DL) approaches, have become promising tools to address nonlinear non-convex problems and high-computation issues, which are mathematically intractable. Particularly, deep Q-network (DQN) has shown its power in solving sophisticated decision-making problems under uncertain and dynamic environments, e.g., human-level game playing~\cite{mnih2013playing, vinyals2019grandmaster} and AlphaGo~\cite{silver2016mastering}. Inspired by the remarkable performance of DQN in various areas, there have been some works exploring its application in wireless communications. DQN provides a principled and robust method to tackle the dynamic environment by making decisions for discrete optimization problems, which bring it the ability to optimize the resource allocation for D2D communications in varying channel state environment. Moreover, as a new technique of DQN, the proposed DDQN provides a more reasonable way to evaluate and execute the action, which avoids the overestimation challenge of legacy DQN algorithms and is more robust to time-varying environment~\cite{vanhasselt2015deep}.

\subsection{Related work}
\subsubsection{Resource allocation in D2D}
The existing works on D2D communications mainly focus on transmit power and channel assignment optimization ~\cite{5910123,7579565,7913583}. In ~\cite{7579565}, EE is maximized by optimizing the transmit power while satisfying the QoS requirement for D2D and cellular users. To overcome the challenges caused by dynamic D2D channels, a co-design of robust spectrum allocation and power optimization has been proposed~\cite{7913583}. While there exists plenty of literature applying optimization tools to solve resource allocation problems for D2D communications, most of them requires intensive computation at the BS to run the optimization algorithms~\cite{8943940}.

As discussed earlier, DL based approaches enable wireless communication users to treat the dynamic environment and make their robust decisions with lower computational complexity. DL has been applied to physical layer processing~\cite{8663966} and resource allocation~\cite{8943940}. Moreover, relying on the local users information and observations, multi-agent reinforcement learning (MARL) based decentralized optimization approaches have been widely applied in wireless communications~\cite{liu2020multiagent,liang2019spectrum,vu2020multiagent,9026965}. Leveraging MARL, D2D users can make their own decisions on transmit power and spectrum sharing policy, which also offloads the computational complexity from the BS to users. In~\cite{liu2020multiagent}, each D2D pair is a learning agent and able to explore the unknown environment. Each D2D user chooses its transmit power level and sub-channel to minimizing long-term system cost. However, the unknown policy and information of other users causes a non-stationary environment. To overcome it, MARL algorithms with improved state observation have been proposed in~\cite{liang2019spectrum,vu2020multiagent} to perform spectrum sharing. Particularly, the D2D links could be deployed as MARL agents, learning to access the channel of cellular users by collectively interacting with the communication environment and receiving the rewards. Such decentralized optimization approach has been verified to achieve the near-optimal performance~\cite{9026965}. Although the above valuable works improve the performance of D2D communications significantly, they mainly focus on the transmit power allocation and channel assignment allocation under fixed communication environment. With the help of RIS, we are able to actively control the communication environment and optimize the resource allocation from a brand new perspective.

\subsubsection{RIS enhanced wireless communications}
Recently, RIS has been explored in a wide range of scenarios, e.g., RIS-enhanced cellular networks beyond 5G, RIS-assisted indoor communications, and IoT applications~\cite{9140329}. Particularly, RIS has been successfully applied in D2D networks in~\cite{cao2020sum,fu2020reconfigurable,ch2020reconfigurable}. Many approaches have been developed to optimize RIS for achieving higher throughput or EE. To solve the non-convex maximizing problems, they tend to find sub-optimal solutions by using the block coordinate descent~\cite{cao2020sum} and Riemannian pursuit method~\cite{fu2020reconfigurable}. To achieve a performance-complexity tradeoff, the projected sub-gradient method is adopted for the phase shift~\cite{ch2020reconfigurable}. However, to enhance the overall system performance, the optimization of RIS becomes a challenge due to the huge number of reflecting elements to optimize~\cite{El_Mossallamy_2020}. The time-varying D2D channel also brings high complexity to optimization algorithms. 

A well-trained ML model is an effective approach to lower computational cost. Although the ML model requires more computations at the training stage, it could be trained offline and is robust to dynamic environment. As a novel branch of ML, DL has been applied for channel estimation and phase shift optimization in RIS-aided communications~\cite{gacanin2020wireless}. Motivated by the applications of DL in solving sophisticated optimization problems, the authors in \cite{taha2019enabling} have applied the DL method for estimating the channels and configuring of RIS. Moreover, DQN has shown its potential in various communication scenarios, e.g., massive multiple input multiple output (MIMO) systems~\cite{huang2020reconfigurable}, RIS-aided unmanned aerial vehicle (UAV)~\cite{liu2020machine}, and non-orthogonal multiple access (NOMA)~\cite{9174801}.

\subsection{Motivation and Contribution}
We consider a reconfigurable intelligent surface enhanced D2D communications system to actively optimize the performance of communication. The challenges occur in several aspects. The fast channel variations of D2D communications makes the conventional resource allocation approaches based on stable channel state information (CSI) not applicable anymore~\cite{7913583}. Most of the current works for the RIS enhanced D2D communication separate the RIS optimization and resource allocation into sub-problems, then leveraging alternating optimization to solve the problem~\cite{yang2020reconfigurable,cao2020sum,fu2020reconfigurable,ch2020reconfigurable}. The large number of RIS elements and alternating optimization means high computational complexity. Additionally, all of these works only consider the phase shift, i.e., passive beamforming design, while neglecting deployment position of the RIS. To overcome the challenges, a low complexity approach is required to achieve real-time optimization\cite{gacanin2020wireless}.

In this paper, we apply a DDQN based structure to optimize the transmit power of D2D users, the channel assignment for D2D pairs, the RIS position, and the phase shift. The major contributions of this paper are summarized as follows.
\begin{enumerate}[1)]

\item The position and phase shift of RIS are jointly optimized at the BS, while the decision-making of channel assignment for D2D links is performed decentrally based on their observation, thus offloading the computational pressure at the BS and enhancing data privacy.

\item A novel reinforcement learning structure is proposed to execute the resource allocation, the phase shift, and RIS position deployment task with lower computational complexity and communication cost. To enhance the robustness and effectiveness of the proposed algorithm, a DDQN algorithm is applied to overcome the overestimation problem.

\item Based on numerical results, the proposed decentralized DDQN architectures can achieve near-optimal performance with low complexity and high robustness.
\end{enumerate}

The rest of this paper is organized as follows. The system
model of RIS enhanced D2D communication system is presented in Section \uppercase\expandafter{\romannumeral2}. Then, the decentralized resource allocation structure and the DDQN based centralized RIS optimization are introduced in Section \uppercase\expandafter{\romannumeral3} and Section \uppercase\expandafter{\romannumeral4}, respectively. Simulation results are presented in Section \uppercase\expandafter{\romannumeral5}. Finally, conclusions are drawn in Section \uppercase\expandafter{\romannumeral6}.

\section{System Model}
In this Section, a RIS enhanced D2D network model is described and an uplink rate maximization problem is formulated. 
\subsection{System settings}
\begin{figure}[t]
\centering
\includegraphics[width=\columnwidth]{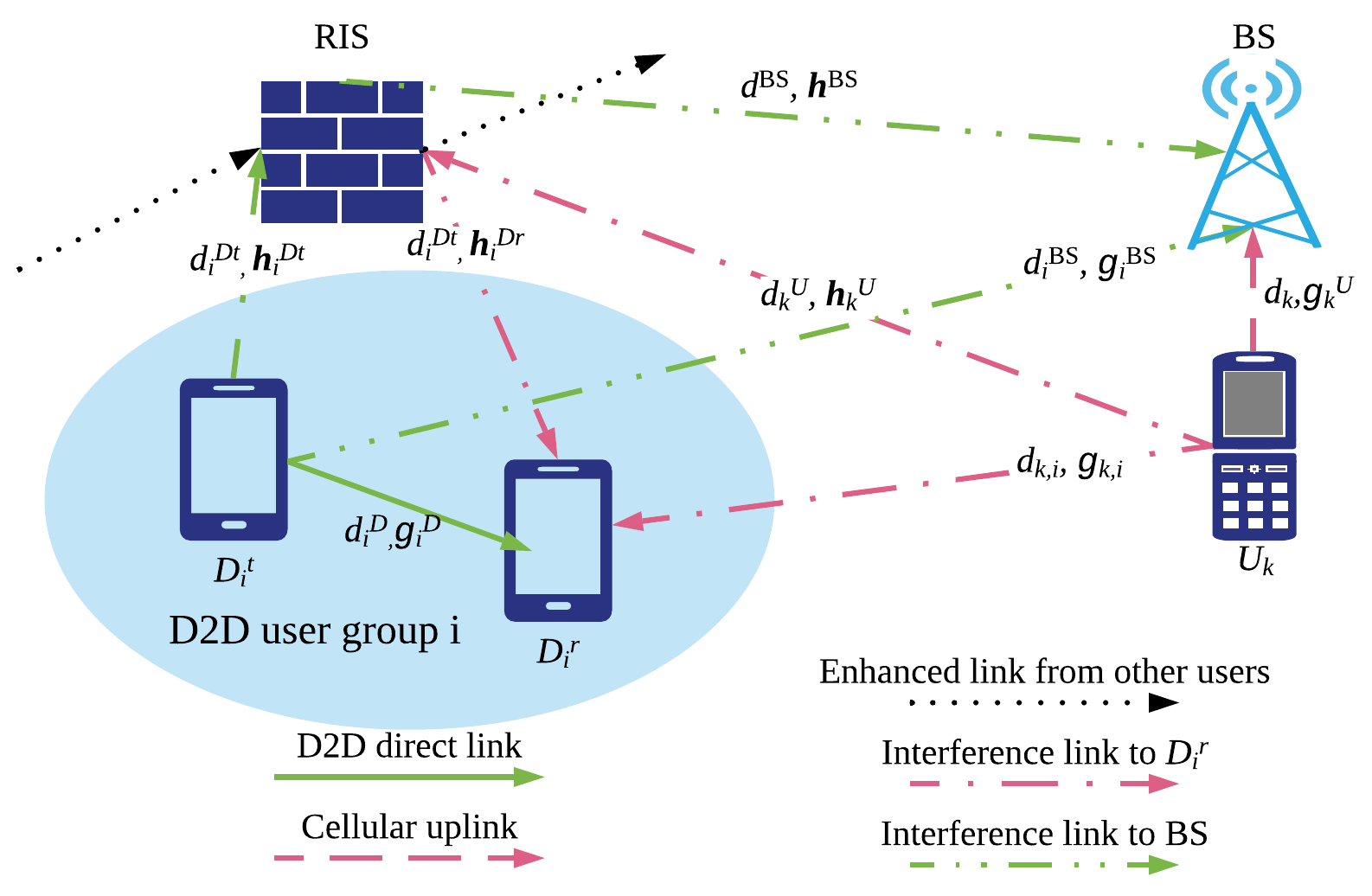}
\centering
\caption{System model of the of RIS enhanced D2D network.}
\label{fig1}
\end{figure}
We consider the uplink transmission in a cellular network, which includes $K$ cellular users communicate with the BS in the cellular mode, and $I$ D2D pairs communicating with each other by sharing the resource blocks (RB) with cellular users. Assuming that the $i$-th D2D transmitter, $D^t_{i}$, communicating with the corresponding receiver, $D^r_{i}$, by reusing the RB assigned to the $k$-th cellular user, $U_k$, then $D^t_{i}$ becomes the source of interference for $U_k$. To enhance the transmission performance of the network, a RIS composed of $N$ passive elements is deployed.

As shown in Fig. \ref{fig1}, we consider that all channels consist of large-scale fading and small-scale fading. The direct links, i.e., the links without the aid of the RIS, are modeled by the Rayleigh fading. The channel gain $g^D_i[k]$ between the D2D pair ($D^t_{i}$ and $D^r_{i}$) over the $k$ sub-band, which is preoccupied by the $k$-th cellular user $U_k$ can be denoted as
\begin{equation}
g^D_i[k] = \sqrt{L(d^D_i)} f^{NLoS}[k],
\label{eq1}
\end{equation}
where $f^{NLoS}[k]$ is a frequency dependent variable that represents small-scale Rayleigh fading power component distribution for $k$-th sub-band. Meanwhile, the large-scale path loss is assumed to be frequency independent and is modeled as $L(d^D_i) = h_0(\frac{d^D_i}{d_0})^{-\alpha}$, where $h_0$ is the path loss constant, $d^D_i$ is the distance between $D^t_{i}$ and $D^r_{i}$, and $\alpha$ represents the path loss exponent. Similarly as (\ref{eq1}), the channel gains for the direct link between $U_k$ and $D^r_{i}$, the link between $U_k$ and the BS, the link between $D^t_{i}$ and the BS over $k$-th sub-band are denoted by $g_{k,i}[k]$, $g^U_k[k]$, and $g^{BS}_i[k]$. 

Due to the existence of line-of-sight (LoS) links, we model the RIS aided links as Rician fading channels. The channel gain $h[n,k]$ of the RIS aided link between users to the $n$-th RIS elements is denoted as
\begin{equation}
h[n,k] = \sqrt{L(d)} (\sqrt{\frac{v}{1+v}}f^{LoS}[k]+\sqrt{\frac{1}{1+v}}f^{NLoS}[k]),
\label{eq2}
\end{equation}
where $v$ is Rician factor and $f^{LoS}[k]$ refers to Rician fading  power  component distribution for the $k$-th sub-band. We define the channel gain $\boldsymbol {h}^{D_t}_i[k]$ for the link between the $i$-th D2D transmitter $D^t_{i}$ and the RIS as
\begin{equation}
(\boldsymbol {h}^{D_t}_i[k])^\mathrm{T} = [h^{D_t}_{i}[1,k], \dots, h^{D_t}_{i}[n,k], \dots, h^{D_t}_{i}[N,k]]^\mathrm{T}, 
\label{eq3}
\end{equation}
where $\boldsymbol {h}(\cdot)^\mathrm{T}$ represent transpose matrix of $\boldsymbol {h}(\cdot)$, $h^{D_t}_{i}[n,k]$ could be defined as (\ref{eq2}), representing the channel gain between $D^t_{i}$ and the $n$-th RIS element. Similarly, the channel gains for the link between the RIS and $D^r_{i}$, the link between $U_k$ and RIS, the link between RIS and BS are defined as $\boldsymbol {h}^{D_r}_i[k] \in {\mathbb C}^{1\times N}$, $\boldsymbol {h}_k^\mathrm{U}\in {\mathbb C}^{N\times1}$ and $\boldsymbol {h}^{BS}\in {\mathbb C}^{1\times N}$ as (\ref{eq3}), respectively. The phase shift and amplitude attenuation $A$ for all the RIS elements can be expressed as $\boldsymbol {\Theta} \triangleq diag[A e^{j{\theta_1}}, A e^{j{\theta_2}}, \dots, A e^{j{\theta_N}}]$, where $A \in [0,1]$ and $\theta \in [0,2\pi)$.

Overall, the channel gain over the $k$-th sub-band for the $i$-th D2D link, $h^D_i[k]$, is given by
\begin{equation}
h^D_i[k] = \underbrace{(\boldsymbol {h}^{D_t}_i[k])^\mathrm{T} \boldsymbol {\Theta} \boldsymbol {h}^{D_r}_i[k]}_{\text{Reflection link}}  + g^D_{i}[k].
\label{eq4}
\end{equation}
Similarly, the overall channel gain between
$U_k$ and $D^r_i$, the channel gain between $l$-th D2D transmitter $D^t_l$ and $D^r_i$, the gain between $U_k$ and the BS, the gain between $D^t_{i}$ and the BS can be represented as $h_{k,i}[k]$, $h^D_{l,i}[k]$, $h^U_k[k]$ and $h^{BS}_i[k]$, respectively.
The signal $y_i[k]$ received by $D^r_{i}$ over the $k$-th sub-band is denoted as
\begin{equation}
y_i[k] = \underbrace{h^D_i[k]\cdot x^D_i}_{\text{Desired signal}} + \underbrace{h_{k,i}[k]\cdot x^U_k}_{\text{Interference signal}} + \underbrace{z}_{\text{Noise}},
\label{eq5}
\end{equation}
where $x^D_i \triangleq \sqrt{p^D_i} u^D_i$ and $x^U_k \triangleq \sqrt{p^U_k} u^U_k$ denote the signal from $D^t_{i}$ and $U_k$, $p^D_i$ and $p^U_k$ denote the transmit power of the $D^t_i$ and $U_k$, and $u^D_i$ and $u^U_k$ represent the unit variance entries with zero mean, and $z\thicksim N(0,\sigma^2)$ denotes the AWGN noise signal with mean 0 variance $\sigma^2$. Then, the signal-to-interference-plus-noise ratio (SINR) at $D^r_i$ and the BS for $U_k$ over the $k$-th sub-band can be denoted as
\begin{equation}
\gamma_i^D[k] = \frac{p^D_i |h^D_i[k]|^2}{I_i[k]+\sigma^2}, 
\label{eq6}
\end{equation}
and
\begin{equation}
\gamma_k^U[k] = \frac{p^U_k |h^U_k[k]|^2}{\sum^I_{i=1} \rho_{k,i}p^D_i|h^{BS}_i[k]|^2+\sigma^2},
\label{eq7}
\end{equation}
respectively, where $\rho_{k,i}$ is the resource reuse coefficient of $U_k$ and $i$-th D2D pair, and $\rho_{k,i}=1$ when $i$-th D2D pair reuses the channel assigned to $U_k$. Otherwise, $\rho_{k,i}=0$. Moreover, the interference to $D^r_i$ is given by
\begin{equation}
    I_i[k] = \rho_{k,i} p^U_k |h_{k,i}[k]|^2+\sum^I_{l=1, l\neq i}\rho_{k,l}p^D_{l}|h^D_{l,i}[k]|^2,
    \label{eq8}
\end{equation}
Then, the ergodic capacity for $i$-th D2D pair and for the $k$-th cellular user $U_k$ can be denoted by
\begin{equation}
    C^D_i[k] = \mathbb{E} [B[k]\log_2(1+\gamma^D_i[k])],
    \label{eq9}
\end{equation}
and
\begin{equation}
    C^U_k[k] = \mathbb{E} [B[k]\log_2(1+\gamma^U_k[k])],
    \label{eq10}
\end{equation}
respectively, where $\mathbb{E[\cdot]}$ represents the statistical expectation of $[\cdot]$, representing the expectation of the rate over the small scale fading distribution, $B_k$ is the bandwidth of $k$-th sub-band. The channel capacity of underlaying D2D networks could be expressed by
\begin{equation}
\begin{aligned}
C&=\sum \limits^K_{k=1}(\sum \limits^I_{i=1} \rho_{k,i} C^D_i[k] + C^U_k[k]).
\end{aligned}
\label{eq11}
\end{equation}

\subsection{Problem formulation}
We aims to maximize the long-term sum rate in (\ref{eq6}) by jointly optimize the phase shift, the position of RIS, the resource reuse coefficient $\boldsymbol \rho = [\rho_{1,1},\dots,\rho_{1,I},\dots,\rho_{K,1},\dots,\rho_{K,I}]$, and the transmit power $\boldsymbol {p}^D = [p^D_{1},\dots,p^D_{I}]$ of D2D transmitters. The joint data rate maximization problem can be formulated as
\begin{maxi!}|l|
{\{\boldsymbol {s}^{RIS},\boldsymbol {\Theta},\boldsymbol \rho, \boldsymbol p^D\}}{C}
{\label{eq12}}{\text{P1:}}
\addConstraint{\gamma_i^D \geq \gamma_{min}^D, \forall i\label{objective:c1} }
\addConstraint{\gamma_k^U \geq \gamma_{min}^U, \forall k\label{objective:c2} }
\addConstraint{\sum^K_{k=1} \rho_{k,i} \leq 1 \label{objective:c3} }
\addConstraint{0 < \theta_n \leq \pi, \forall n \in N\label{objective:c4} }
\addConstraint{\boldsymbol {s}^{RIS}\in \mathbb{R}^2, \label{objective:c5} }
\end{maxi!}
where $\gamma_{min}^D$ and $\gamma_{min}^U$ are the minimum SINR requirements at the D2D receiver and the BS, respectively. Coordinate $\boldsymbol {s}^{RIS}$ restricts a two dimensional (2D) space for the installation of the RIS. Constraint (\ref{objective:c3}) assumes that each D2D pair only occupies one RB. Due to hardware limitations, RIS elements can only provide discrete phase shifts. This constraint and (\ref{objective:c3}) make (P1) non-convex. To solve the non-convex problem, we have to utilize exhaustive search, which is impractical when the number of D2D pairs, cellular users, and the number of RIS elements become large. Generally, classical optimization tools can be leveraged to acquire suboptimal solutions~\cite{cao2020sum,fu2020reconfigurable,ch2020reconfigurable}. Alternatively, instead of solving challenging non-convex problem by mathematical tools, we leverage an deep RL based algorithm, which is more applicable to solve problems with high dimension inputs as well as large state and action space which will be detailed in Section \uppercase\expandafter{\romannumeral3}.


\section{Resource allocation optimization by multi-agent reinforcement learning}
In this section, we study the transmit power optimization and channel assignment optimization for each D2D pair. The optimization objective is to jointly optimize the resource allocation for D2D pairs, plus the position and phase shift for the RIS. Instead of optimizing the configuration of RIS and the resource sharing centrally at the BS, we propose the decentralized resource sharing scheme and centralized RIS optimization approach. By decoupling the joint optimization into sub-problems, we not only reduce the computational pressure of the BS significantly, but also enable the D2D users to determine their resource sharing policies by local information. 

As shown in Fig. \ref{fig1}, D2D pairs reuse the RB occupied by cellular networks. Given an arbitrary RIS implementation, i.e., the position and phase shift of the RIS, the resource allocation optimization problem can be simplified into
\begin{maxi!}|l|
{\{\boldsymbol \rho, \boldsymbol {p}^D\}}{C}
{\label{eq13}}{\text{P2:}}
\addConstraint{\gamma_i^D \geq \gamma_{min}^D, \forall i\label{objective:c6} }
\addConstraint{\gamma_k^U \geq \gamma_{min}^U, \forall k\label{objective:c7} }
\addConstraint{\sum^K_{k=1} \rho_{k,i} \leq 1. \label{objective:c8} }
\end{maxi!}

\subsection{System description}
Generally, the resource allocation optimization problem can be modeled as a linear sum assignment programming (LSAP) problem and can be solved by Hungarian algorithm~\cite{3800020109} with computational complexity $O(n^3)$. The complexity is much higher if we take the transmit power of D2D users into account. The high complexity of the Hungarian algorithm makes real-time optimization is impractical in the proposed D2D communications scenario. Additionally, the algorithm needs to be robust for fast channel variations and unstable CSI for different RIS implementations. Leveraging DDQN, we can train the agents under different CSI conditions so that it can be adapted to the various communications system. 

Then the channel assignment and transmit power optimization problem can be modeled as a MARL problem. It is noted that the DDQNs at the D2D users is should be trained offline. Actually, updating the resource allocation policy too quickly can cause challenges on convergence performance when we train the DDQN for RIS optimization. This is because even if the RIS controller takes the exactly same action, the rewards would be various for different resource allocation policies, making the algorithm hard to converge. The unstable reward requires a robust resource allocation algorithm so that it can works under different RIS implementations. 

\subsection{Concept of Reinforcement learning and deep Q-network}
\begin{figure}[t]
\centering
\includegraphics[width=\columnwidth]{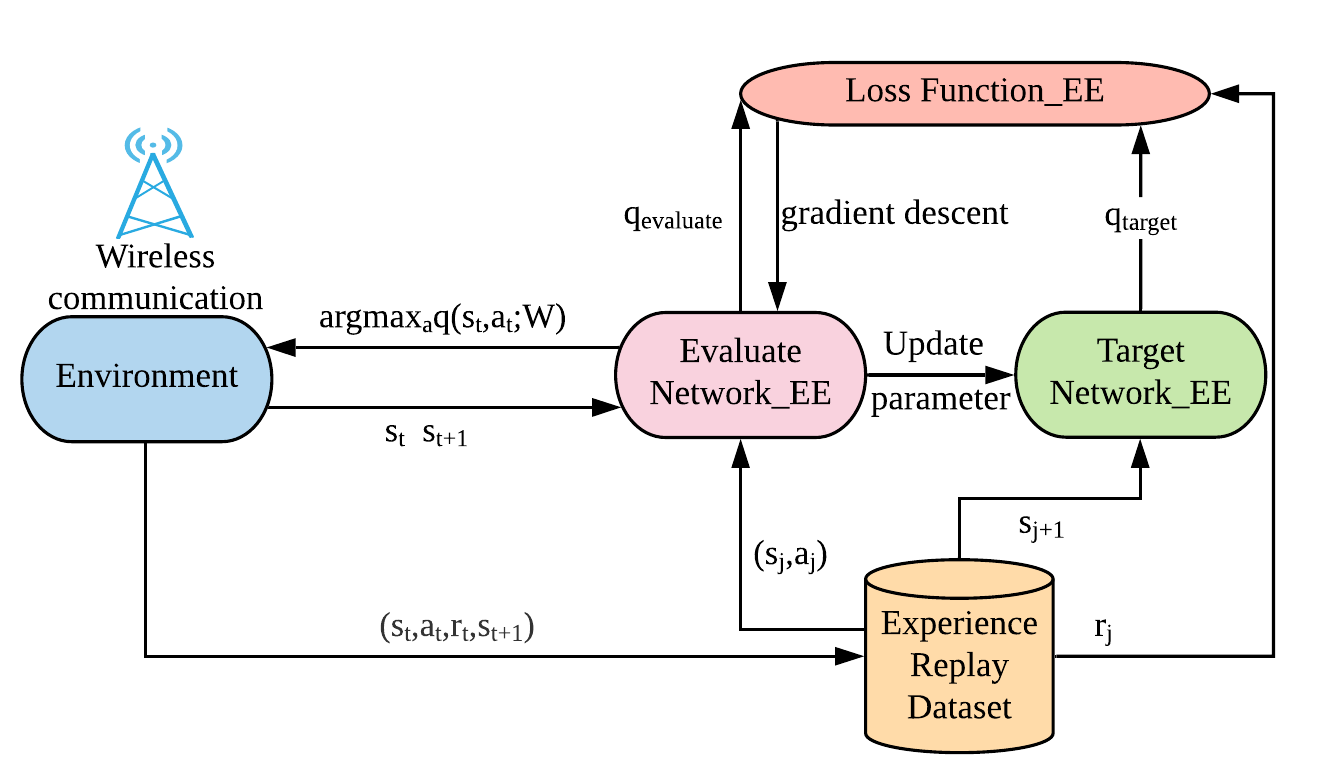}
\centering
\caption{The interaction of the DQN with the environment.}
\label{fig3}
\end{figure}

RL is a branch of ML paradigm that allows agents to learn the optimal policy by the trial-and-error interaction with the environment to maximize the desired reward. Mathematically, the RL can be modeled as an markov decision process (MDP), including environment state $\cal S$, actions $\cal A$, and the reward $\cal R$ which can be determined for each state-action pair. During each time slot $t$, each agent observes the state $s_t \in \cal S$ and then take an action $a_t \in \cal A$ according to a certain policy $\pi$. Then the agent receives the corresponding reward $r_t$ and turn to the next state $s_{t+1}$, which is determined by action $a_t$ but independent of the past states. Formally, this process can be denoted by a transition tuple $e_t = (s_t,a_t,r_t,s_{t+1})$. The interaction process is shown in Fig.~\ref{fig3}. 

During each time slot $t$, the objective of RL is to maximize the cumulative desired return from time $t$ to the future, which can be expressed by 
\begin{equation}
R_t = \sum^{\infty}_{\tau=0}\gamma^\tau r_{t+\tau},
\label{eq14}
\end{equation}
where $\gamma \in (0,1)$ represents the discount factor which represents the impact of the future reward. The expectation reward for a state-action pair $(s,a)$, the action-value function, is defined as
\begin{equation}
q^\pi(s,a) = \mathbb{E}_\pi[R_t|s_t=s, a_t=a],\
\label{eq15}
\end{equation}
where policy $\pi$ is defined as a mapping from state $\cal S$ to the probability of choosing each action in $\cal A$. 

The objective of RL is to find a optimal policy $\pi^* = \arg\max_{\pi} q^\pi(s,a)$. The optimal action-value function obeys an important identity known as the Bellman equation. The optimal policy is to select the action that maximizes the expected Q-value~\cite{mnih2013playing}:
\begin{equation}
q^*(s_t, a_t)=\mathbb{E}[r_t+\gamma\max\limits_{a'\in {\cal A}} q^*(s_{t+1},a')|s_t,a_t].
\label{eq16}
\end{equation}
Authors in \cite{sutton2018reinforcement} have shown that, $q(s_t, a_t) \rightarrow q^*(s_t, a_t)$ as $t \rightarrow \infty$. However, it is impractical since the iteration is discrete. Instead, the NNs are applied to be function approximator to estimate the action-value function, i.e., $q(s_t, a_t; \boldsymbol W) \approx q^*(s_t, a_t)$, which is the basic idea of the DQN. When the state and action space become large, this method does not need to maintain the large Q-table as conventional RL approaches do, thereby expanding the applications of RL in wireless communications greatly.

The training data set, also named replay memory ${\cal D}=[\boldsymbol e_1,\dots, \boldsymbol e_t, \dots]$ for NN is stored according to agent's experience at each iteration $t$, where the experience $\boldsymbol e_t=(s_t,a_t,r_t,s_{t+1})$ is called transition, including the state, action and reward information. The training minibatch $(s_j,a_j,r_j,s_{j+1})$ is sampled from the training data set. During the training process, parameters are updated to the Q estimation network at each step to generate the estimated Q-value. Q target network is updated after every $g$ steps according to the parameters in the Q estimation network. The training process for DDQN is to minimize the error function which represents the estimated Q-value and the realistic Q-value. For the DDQN in this work, the error function can be expressed by:
\begin{equation}
\text{Loss}(\boldsymbol W)=\mathbb E[(q_{target}-q(s_j,a_j;\boldsymbol{W}))^2],
\label{eq17}
\end{equation}
where $q_{target}=r_j+\gamma \max_{a'}q(s_{j+1},a';\boldsymbol W^-)$ is the target Q-value for minibatch $j$, which is the output of Q target network, $\boldsymbol W$ and $\boldsymbol W^-$ denotes the weights of the evaluation network and the target network, respectively. The weights are optimized by the gradient descent method~\cite{mnih2013playing}. 

\subsection{Double DQN algorithm}
The DQN algorithm can achieve a near-optimal performance in some scenarios, while sometimes it causes the overestimate problem. The target Q-value is approximately generated by the target network by maximizing the action-value function, while this target value is even higher than the true optimal action-value. The overestimate problem is severest when the number of actions becomes large, affecting the convergence and performance of the learned strategies. The idea of DDQN is decomposing action selection and action evaluation~\cite{vanhasselt2015deep} to reduce overestimations. Unlike DQN that uses the evaluate network to estimate the action-value function and select action at the same time, DDQN uses the target network when evaluating the action-value function. In other words, the DDQN uses the evaluate network to select the action, while using the target network to fairly evaluate this action. The updated target Q-value function in DDQN is defined as
\begin{equation}
q_{target}=r_j+\gamma q(s_{j+1},\mathop{\arg\max}\limits_{a'\in {\cal A}} q(s_{j+1},a';\boldsymbol W);\boldsymbol W^-).
\label{eq23}
\end{equation}

Note that DDQN is a model-free algorithm, which guarantees its robustness for different scenarios. Meanwhile, it is an off-policy algorithm which learns from the greedy policy and choose the action according to $\epsilon$-greedy algorithm to make a trade-off between exploitation and exploration. The agent will choose actions uniformly from ${\cal A}$ with a probability of $\epsilon$, while choosing the action  $a = \max_a q(s,a;W)$ which maximizes the Q-value with a probability of (1-$\epsilon$). In this paper, we leverage an improved algorithm called decaying $\epsilon$-greedy algorithm as shown in~\cite{liu2020ris}, so that we can achieve a better explored performance at the beginning and converged performance in the end.

\subsection{Multi-agent observation state}
In the multi-agent resource allocation process, each D2D link is modeled as an agent, concurrently making there own decision based on the local observation. Given the current environment state $s_t$, agent $D_i$ generates the unique observation $z^{(i)}_t$ from $s_t$ at each time slot $t$, according to the observation function $z^{(i)}_t = o(s_t,i)$. Then it takes an action $a^{(i)}_t$, forming the joint action $\boldsymbol a_t$ with all the other agents. Then agents will receive an reward $r_{t}$ and the environment turn to the next state $s_{t+1}$. Observations $z^{(i)}_{t+1}$ in the next time slot will then be generated agent $D_i$. 

An agent cannot acquire the global environment state $s_t$ which contains the global channel information and agents behaviour directly, thus the state design in~\cite{9247965} based on global SINR information is not applicable. Rather than the position information based state definition, the CSI based state definition enhances the robustness of the model. In other words, for the $i$-th D2D receiver, the observation space includes: (\romannumeral1) local channel information $h^D_i[k]$; (\romannumeral2) the interference channel from other D2D transmitters $h^{D}_{l,i}[k]$ for $l \neq i, l \in I$; (\romannumeral3) the interference channel to the BS $h^{BS}_{i}[k]$; (\romannumeral4) Interference from cellular users $h_{k,i}[k]$ for $k \in K$; (\romannumeral5) Interference power $I_i[k]$. The information of channel (\romannumeral1), (\romannumeral2) and (\romannumeral4) can be estimated by D2D receiver accurately, while (\romannumeral3) can be estimated and broadcast by the BS. Additionally, interference power (\romannumeral5) can be measured by D2D receiver. Thus, the observation space of $i$-th agent at time $t$ can be denoted by $o(s_t,i) = \{\{H_i[k]\}_{\forall k \in K},\{I_i[k]\}_{\forall k \in K}\}$, where $H_i[k] = \{h^D_i[k], \{h^D_{l,i}[k]\}_{\forall l \in I, l \neq i}, h^{BS}_{i}[k], h_{k,i}[k]\}$.

Particularly, the multi-agent learning process can be described as Markov game. The state transition depends on actions taken by all of the agents, i.e., the joint action contributes to the state shift. Apart from the action taken by an agent itself, the actions of other agents can impact the reward of the agent, forming an unstable environment. The nonstationary environment from the view of each agent leads to nonstationary Q-function, making RL hard to converge. The nonstationarity challenge is tackled in~\cite{gundogan2020distributed} with a unique state, which includes view-based positional distribution and shared position information by each vehicle.

The problem is severest when combining with deep learning. DDQN uses experience replay to feed the neural network, while the environment that generated the data in the agent’s replay memory is different from the current environment, and the convergence of the learning process is affected. To enable replay memory in MARL, authors in~\cite{foerster2017stabilising} designed a low dimensional \textit{fingerprint} which includes the information of policy change of other agents. The policy change is highly correlated with iteration times $e$ and the exploration rate $\epsilon$. In other words, the observation space for $i$-th agent can be expressed as
\begin{equation}
z^{(i)}_t = \{o(s_t,i),e,\epsilon\}.
\label{eq18}
\end{equation}
Such fingerprint allows an agent to expect the policy change of other agents, thus improving the stationarity of the environment.

\subsection{Actions and rewards definition}
Assuming that orthogonal frequency division multiplexing (OFDM) is applied for the uplink of cellular network, which means cellular users communicate with the BS on disjoint RBs. Each D2D pair can choose one of $K$ RB that is preoccupied with a cellular user. The range of D2D transmit power including $A_p$ multiple discrete levels is $[0, p^D_{max}]$. As the result, the dimension of the action space is equal to $A_p \times K$. The actions of all agents form a joint action $\boldsymbol{a}_t$ which represents the resource reuse scheme.



Reward represents the objective of the optimization. All agents receive the same reward $r_t$ according to the joint action $\boldsymbol{a}_t$ such that encouraging cooperative behaviors. The reward can be defined as
\begin{equation}
r_t=
\begin{cases}
C,&\text{if (\ref{objective:c6}) and (\ref{objective:c7}) are satisfied};\\
0,&\text{else}.
\end{cases}
\label{eq19}
\end{equation}

\begin{algorithm}[t]
\caption{MARL algorithm for the resource allocation.}
\label{algo1}
\begin{algorithmic}[1]
\STATE \textbf{Input}: Start environment simulator, initialize D2D devices, cellular users and the BS;\
\STATE Initialize the DDQN for each D2D pair;\
\FOR{each episode}
\STATE Initialize the implementation of RIS randomly;\
\STATE Update the large-scale fading channel;\
\FOR{each time slot $t$}
\FOR{each D2D agent $i$}
\STATE Observe $z^{(i)}_t$;\
\STATE Choose action $a^{(i)}_t$ according to the observation $z^{(i)}_t$ and $\epsilon$-greedy algorithm;\
\ENDFOR
\STATE Form the joint action $\boldsymbol{a}_t$ and receive reward $r_{t}$;\
\STATE Update the small-scale fading channel;\
\FOR{each D2D agent $i$}
\STATE Observe $z^{(i)}_{t+1}$;\
\STATE Store transition $e_t = (z^{(i)}_t,a^{(i)}_t,r_{t},z^{(i)}_{t+1})$ in ${\cal D}_i$;\
\ENDFOR
\ENDFOR
\FOR{each D2D agent $i$}
\STATE Replay memory:\
\STATE Sample random minibatch of transitions\\ $e_j = (z^{(i)}_j,a^{(i)}_t,r_{j},z^{(i)}_{j+1})$ in ${\cal D}_i$;\
\STATE Calculate $q_{target}$ by (\ref{eq23})\
\STATE Perform a gradient descent step on \\ $(q_{target}-Q(z^{(i)}_j,a^{(i)}_t;\boldsymbol{W}))^2$;\
\ENDFOR
\ENDFOR
\end{algorithmic}
\end{algorithm}

\subsection{Training algorithm}
As introduced above, each DDQN at the D2D pairs takes its observation state as the input. Several fully connected layers are leveraged as the hidden layer. During the training and testing phases, the RIS is randomly implemented and updated at the beginning of each episode. One training episode contains several time slots during which the agents interact with the wireless communication environment and store the experience in the training data sets, i.e., replay memories. The details of the training algorithm for the MARL at the D2D pairs is shown in \textbf{Algorithm \ref{algo1}}.
\section{RIS optimization process}
\begin{figure}[t]
\centering
\includegraphics[width=\columnwidth]{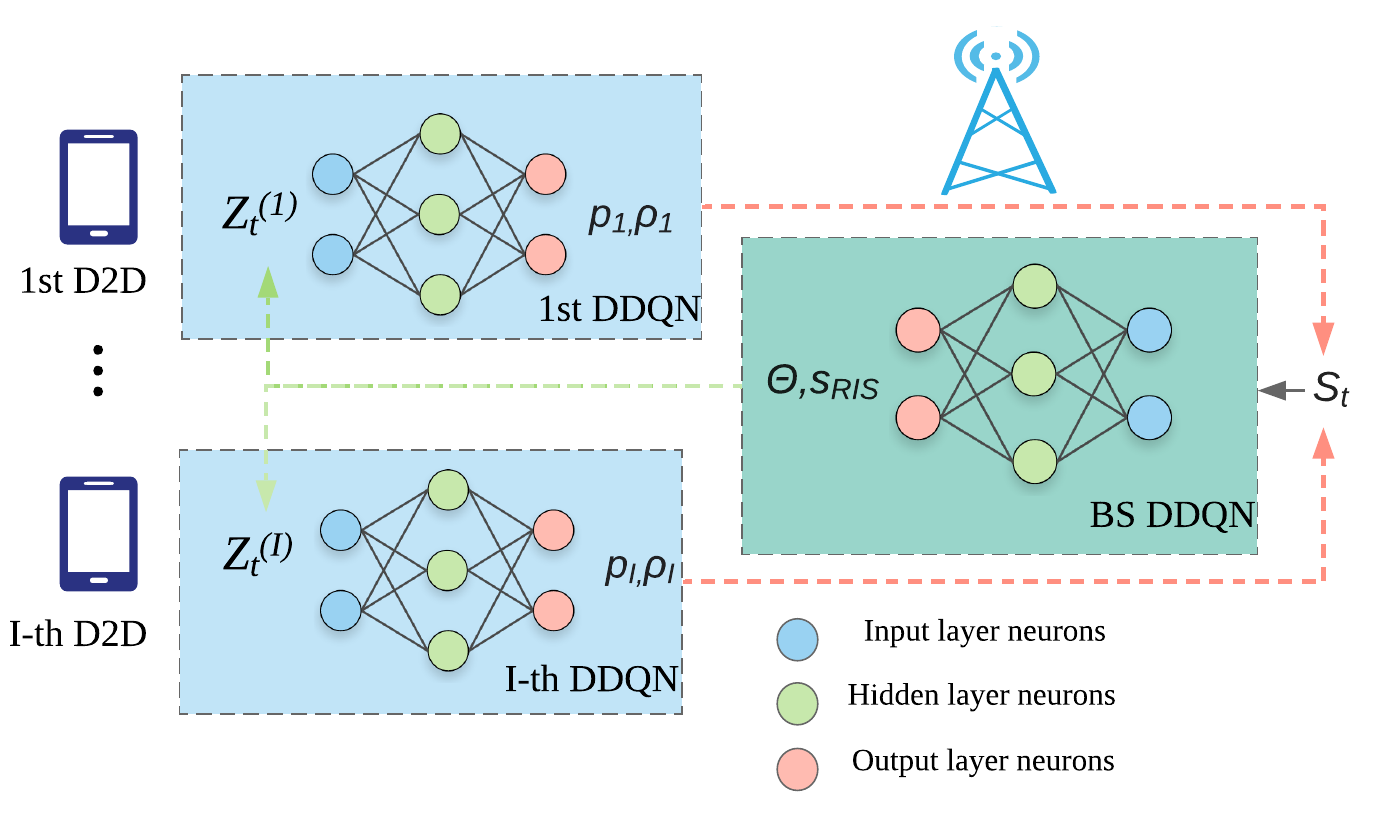}
\centering
\caption{Architecture of the proposed DDQN algorithm.}
\label{fig2}
\end{figure}

After resource allocation decisions are made by D2D pairs, the resource sharing scheme will be sent to the BS as a part of the input information of the DDQN to optimize the RIS. The optimized RIS position and phase shift information will be broadcast to each D2D pair as shown in Fig. \ref{fig2}. Based on the resource sharing information, the sum rate maximization problem can be formulated as
\begin{maxi!}|l|
{\{\boldsymbol {\Theta},\boldsymbol {s}^{RIS}\}}{C}
{\label{eq20}}{\text{P3:}}
\addConstraint{\gamma_i^D \geq \gamma_{min}^D, \forall i\label{objective:c9} }
\addConstraint{\gamma_k^U \geq \gamma_{min}^U, \forall k\label{objective:c10} }
\addConstraint{0 < \theta_n \leq \pi, \forall n \in N\label{objective:c11} }
\addConstraint{\boldsymbol {s}^{RIS}\in \mathbb{R}^2, \label{objective:c12} }
\end{maxi!}
Based on the resource allocation information, a DDQN learning model is proposed at the base station to solve the joint RIS positioning and phase shift problem. Particularly, the RL components are first defined and the DDQN components are then introduced. The proposed algorithm is now explained in detail.

\subsection{RL components definition}
In our proposed algorithm, the state contains the position information of D2D users, cellular users, BS and RIS, the resource sharing coefficient $\boldsymbol{\rho}$, the transmit power of each D2D transmitters, as well as the phase shift $\boldsymbol \Theta$. The position vectors $\boldsymbol{S}^{D_t}=[\boldsymbol{s}^{D_t}_1,\dots,\boldsymbol{s}^{D_t}_I]$, $\boldsymbol{S}^{D_r}=[\boldsymbol{s}^{D_r}_1,\dots,\boldsymbol{s}^{D_r}_I]$, $\boldsymbol{S}^U=[\boldsymbol{s}^U_1,\dots,\boldsymbol{s}^{U}_K]$, $\boldsymbol{s}^{RIS}$ and $\boldsymbol{s}^{BS}$ represents the position information of D2D transmitters, D2D receivers, cellular users, RIS and BS, respectively. The input state ${\cal {S}} = [\boldsymbol{S}^{D_t},\boldsymbol{S}^{D_r},\boldsymbol{S}^U,\boldsymbol{s}^{RIS},\boldsymbol{s}^{BS}; \boldsymbol \rho;\boldsymbol {p}^D;\boldsymbol \Theta]$, which has a cardinality $|\cal S|$ of ($3I + K + N + 2$).

Action set $\cal A$ represents the possible action choice for the RIS controller. Generally, the position of RIS are fixed after installation, while the phase shift can be adjusted, so the action space contains the phase shift adjustment and position choice of RIS. At iteration $t$, action $a_t$ consists of two parts: \romannumeral1) the variable quantity of phase shift matrix, $\Delta \boldsymbol \Theta = \{\Delta \theta_1, \dots, \Delta \theta_N\}$, where $\Delta \theta_n \in \{-\delta, 0, +\delta \}, \forall n \in N$; \romannumeral2) the position choice of RIS, $\boldsymbol s \in \{v_1, \dots, v_O\}$, where $O$ represents the number of grids in the communications system. Formally, the action $a_t = [\Delta \boldsymbol \Theta; \boldsymbol {s}^{RIS}]$, which has a cardinality $|a|$ of ($N + 1$). Action set $\cal A$ includes all possible actions with the cardinality $|{\cal A}| = 3^N\times O$.

The reward represents whether we encourage or punish an action, so it is defined based on the objective function given in (\ref{eq6}). For a successful transmission at iteration $t$, i.e., the constraints (\ref{objective:c1}) and (\ref{objective:c2}) are satisfied, the reward $r_\text{s}$ can be defined as $r_\text{s}(t) = C(t)$, where $C(t)$ represents the achievable rate $C$ at iteration $t$. However, if any of the constraints are not satisfied, the expected QoS cannot be achieved. This kind of action results in a penalty due to energy waste, and we defined the new reward for the transmission failure as
\begin{equation}
r_\text{f}(t) = 
\begin{cases}
\sum \limits^K_{k=1} \sum \limits^I_{i=1} \rho_{k,i} C^D_i[k],&\text{if (\ref{objective:c1}) is not satisfied};\\
\sum \limits^K_{k=1} C^U_k[k],&\text{if (\ref{objective:c2}) is not satisfied};\\
0,&\text{otherwise};
\end{cases}
\label{eq21}
\end{equation}
The fail reward is to encourage the communication system to improve the SINR which is not satisfied with the requirement.
The overall reward can be expressed as
\begin{equation}
r_t=
\begin{cases}
r_\text{s}(t),&\text{if (\ref{objective:c1}) and (\ref{objective:c2}) are satisfied};\\
r_\text{f}(t),&\text{else}.
\end{cases}
\label{eq22}
\end{equation}

\subsection{Proposed double DQN algorithm for the control of RIS}
Leveraging the NN, the DDQN model can find the relationship between the input location information and the corresponding optimized deployment of RIS. The components in DDQN is defined as
\begin{itemize}
\item Agent: The agent in our DDQN model is the BS. The BS will process the inputs and execute the outputs of DDQN to control RIS.
\item Input: The DDQN model takes the states $\cal S$ as the input, which includes the position and phase shift information.
\item Output: The output of the DDQN model is the evaluated Q-value for state-action pairs. The output layer contains $|{\cal A}|$ units, which represents the number of possible actions. As shown in Fig. \ref{fig3}, two identical networks are set: evaluation network and target network. In the evaluation network, the current state $s_j$ is the input information, and the output is the evaluate Q-value for each action. In the target network, the next expected state $s_{j+1}$ is the input, while the output is the Q-value for each action in the next state.
\end{itemize}
The training process for the DDQN is shown in Section \uppercase\expandafter{\romannumeral3}. By receiving the input information of position information of D2D pairs, cellular users, and the BS from the wireless environment, the RIS controller can train the weights and update NNs to estimate the action-value function. The proposed DDQN algorithm for RIS optimization is shown in \textbf{Algorithm \ref{algo2}}.

\begin{algorithm}[t]
\caption{DDQN algorithm for the RIS optimization.}
\label{algo2}
\begin{algorithmic}[1]
\STATE \textbf{Input}: Environment simulator, Q network, replay memory $\cal D$, minibatch size;\
\STATE Initialize: action-value function Q with random weights $\boldsymbol W$, replay memory $\cal D$, RIS position and phase;\
\FOR{each episode}
\STATE Execute multi-agent DQN and perform resource allocation;\
\STATE Update the large-scale fading channel;\
\FOR{each time slot $t$}
\STATE Observe $s_t$;\
\STATE Choose $a_t \in \cal A$ according to $\epsilon$-greedy algorithm;\
\STATE Execute $a_t$ and calculate reward $r_t$ by (\ref{eq10});\
\STATE Update the small-scale fading channel;\
\STATE Observe $s_{t+1}$;\
\STATE Store transition $(s_t,a_t,r_t,s_{t+1})$ in $\cal D$;\
\ENDFOR
\IF{learning begins}
\STATE Replay memory:\
\STATE Sample random minibatch of transitions\\ $(s_j,a_j,r_j,s_{j+1})$ in $\cal D$;\
\STATE Calculate $q_{target}$ according to (\ref{eq23});\
\STATE Perform a gradient descent step on \\ $(q_{target}-q(s_j,a_j;\boldsymbol{W}))^2$;\
\STATE Update the target network every $g$ episodes;\
\ENDIF
\ENDFOR
\STATE \textbf{Return}: action-value function and optimized action $a$.
\end{algorithmic}
\end{algorithm}

\subsection{Computational complexity and communication cost analysis}
Generally, floating point operations (FLOPs) is used to measure the computational complexity. For each fully connected layer, the number of FLOPs is $[N_{in} + (N_{in}-1) + 1]\times N_{out}$, where $N_{in}$ and $N_{out}$ represents the number of neurons. For the DQN at the BS, the number of FLOPs is $\text{FLOPs}(BS) = 2[(2I+k+2)\times N^{BS}_1+N^{BS}_1\times N^{BS}_2+N^{BS}_2\times N^{BS}_3+N^{BS}_3 \times (3^N\times O)]$, where $N^{BS}_\mu$ represents the number of neurons in $\mu$-th layer of the DQN at the BS. For the DQN at each D2D pair, the number of FLOPs is $\text{FLOPs}(i) = 2[N_{in}(i)\times N_1(i)+N_1(i)\times N_2(i)+N_2(i)\times N_{out}(i)]$. Thus, the overall computational complexity can be expressed by

\begin{equation}
\text{FLOPs}=\underbrace{\text{FLOPs}(BS)}_{\text {At the BS}}+\underbrace{\sum^I_{i=1} \text{FLOPs}(i)}_{\text{At D2D pairs}}.
\label{eq24}
\end{equation}

The proposed algorithm could be trained offline because it is robust to the dynamic environment. Compared to the alternative maximization (AM) approaches~\cite{cao2020sum,fu2020reconfigurable,ch2020reconfigurable} that optimize the resource allocation and RIS configuration by iterations, the proposed trained model only requires a little computational complexity to generate solutions.

Compared with the centralized algorithms that the users need to upload the local information to the BS and receive the optimized control signals from the BS, the proposed algorithm enables users to complete the resource allocation process locally, thus reduce the communication cost significantly. In a nutshell, the proposed DDQN based algorithm outperforms other benchmarks in varying wireless communication environments, where the solution generated by AM approaches via previous iterations may not be applicable for the current environment.

\section{Numerical Results}
\addtolength{\topmargin}{0.01 in}
In this section, the performance of the proposed distribute DDQN (D\_DDQN) based algorithm is evaluated by comparing it with the benchmark algorithms. Assuming that the cellular users are uniformly distributed in a $1000\text{m} \times 1000\text{m}$ square. The whole area is divided into $O=25$ identical squares, where RIS can be installed in any of them. The channel models parameters are listed in TABLE \uppercase\expandafter{\romannumeral1}. 


In the proposed MARL structure, each DQN consists of 5-layer fully connected (FC) neural networks with 3 hidden layers. The number of neurons in the three hidden layers are set to 500, 250, and 125, respectively. We apply the rectified linear unit (ReLU) function as the activate function, which is defined as $f(x) = \max(0,x)$, while the RMSProp optimizer~\cite{Ruder16} is applied to train the NNs. For each DQN at D2D users for the resource allocation, the number of training episodes is set as 1500. Note that the trained resource allocation model only needs to be updated when the wireless communication system experience significant changes, thus the resource allocation model is first trained and remains unchanged during the optimization of the RIS. We train the DDQN for RIS optimization for 3000 episodes and the exploration rate $\epsilon$ decreases from 1 to 0.05 over 2000 episodes linearly. The discount factor $\gamma$ is set to 0.9.

We compared the proposed D\_DDQN algorithm with other benchmark algorithms derived from the following schemes.
\begin{enumerate}[1)]

\item \textbf{Baseline 1}: Baseline 1 is achieved by the optimal resource allocation using exhaustive search from all possible channel assignments and RIS implementations.

\item \textbf{C\_DDQN}: In this scheme, a centralized DDQN is applied for the channel assignment and RIS optimization at the central BS. 

\item \textbf{D\_DQN}: The DQN for RIS optimization, while the resource allocation is performed by MARL. This scheme differs from the proposed algorithm through the RIS optimization method.

\item \textbf{Baseline 2}: Baseline 2 shows the performance of the random channel assignments and RIS implementations at each time step. Actions are chosen randomly from the action space $\cal A$ and agents do not learn from the environment.

\item \textbf{Without RIS}: This scheme does not deploy RIS for enhancement, i.e., it just involves resource allocation optimization by MARL.

\end{enumerate}

\begin{table}[t]
\begin{center}
\caption{Simulation Parameters}
\begin{tabular}{|c|c|}
\hline
\textbf{Parameter}&
\textbf{Value} \\ \hline
Number of D2D users $I$&
4\\
\hline
Number of Cellular users (sub-bands) $K$&
4\\
\hline
Path loss exponent &
3\\
\hline
Phase shift variable quantity $\delta$&
$\frac{\pi}{4}$\\
\hline
number of RIS elements $N$&
16\\
\hline
Cellular transmit power range&
$23$dBm\\
\hline
D2D transmit power range&
$[0,24]$dBm\\
\hline
Number of discrete levels $A_p$&
9\\
\hline
Minimum SINR requirements for D2D receiver $\gamma^D_{min}$&
-10dB\\
\hline
Minimum SINR requirements for BS $\gamma^U_{min}$&
-13dB\\
\hline
Carrier frequency&
2GHz\\
\hline
Bandwidth of each sub-band&
1MHz\\
\hline
Cellular antenna height&
1m\\
\hline
D2D antenna height&
1m\\
\hline
BS antenna height&
25m\\
\hline
Bandwidth of each sub-band&
1MHz\\
\hline
Noise power $\sigma^2$&
$-115$dBm\\
\hline
Fast fading update&
Every 1ms\\
\hline
\end{tabular}
\end{center}
\label{tab3}
\end{table}



\subsection{Training Performance of Proposed Algorithms}

\begin{figure}[t]
\centering
\subfigure[Loss comparison of double DQN and conventional DQN.]{
\begin{minipage}[t]{\linewidth}
\centering
\includegraphics[width=0.9\columnwidth]{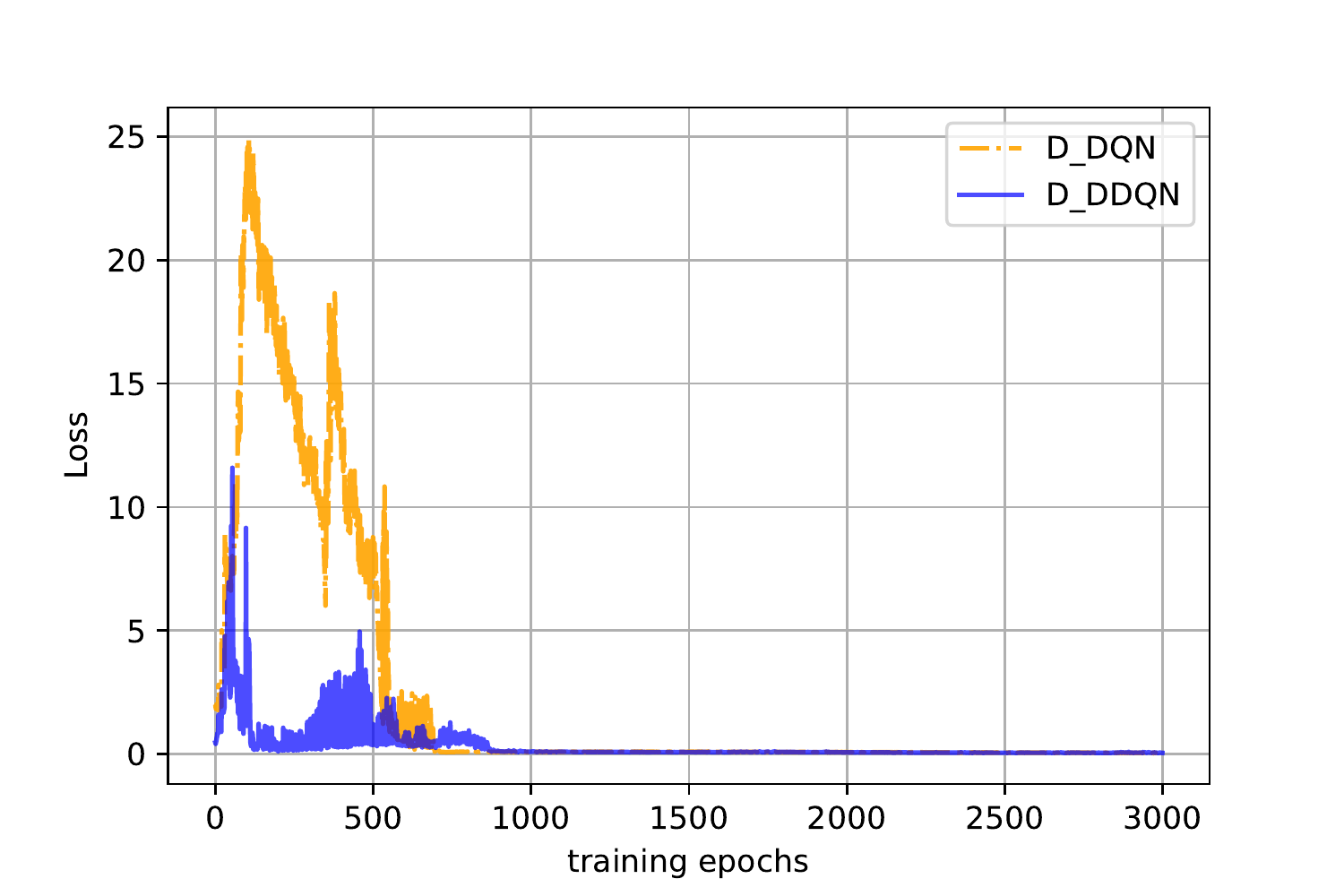}
\end{minipage}%
}%

\subfigure[Reward comparison of double DQN and conventional DQN.]{
\begin{minipage}[t]{\linewidth}
\centering
\includegraphics[width=0.9\columnwidth]{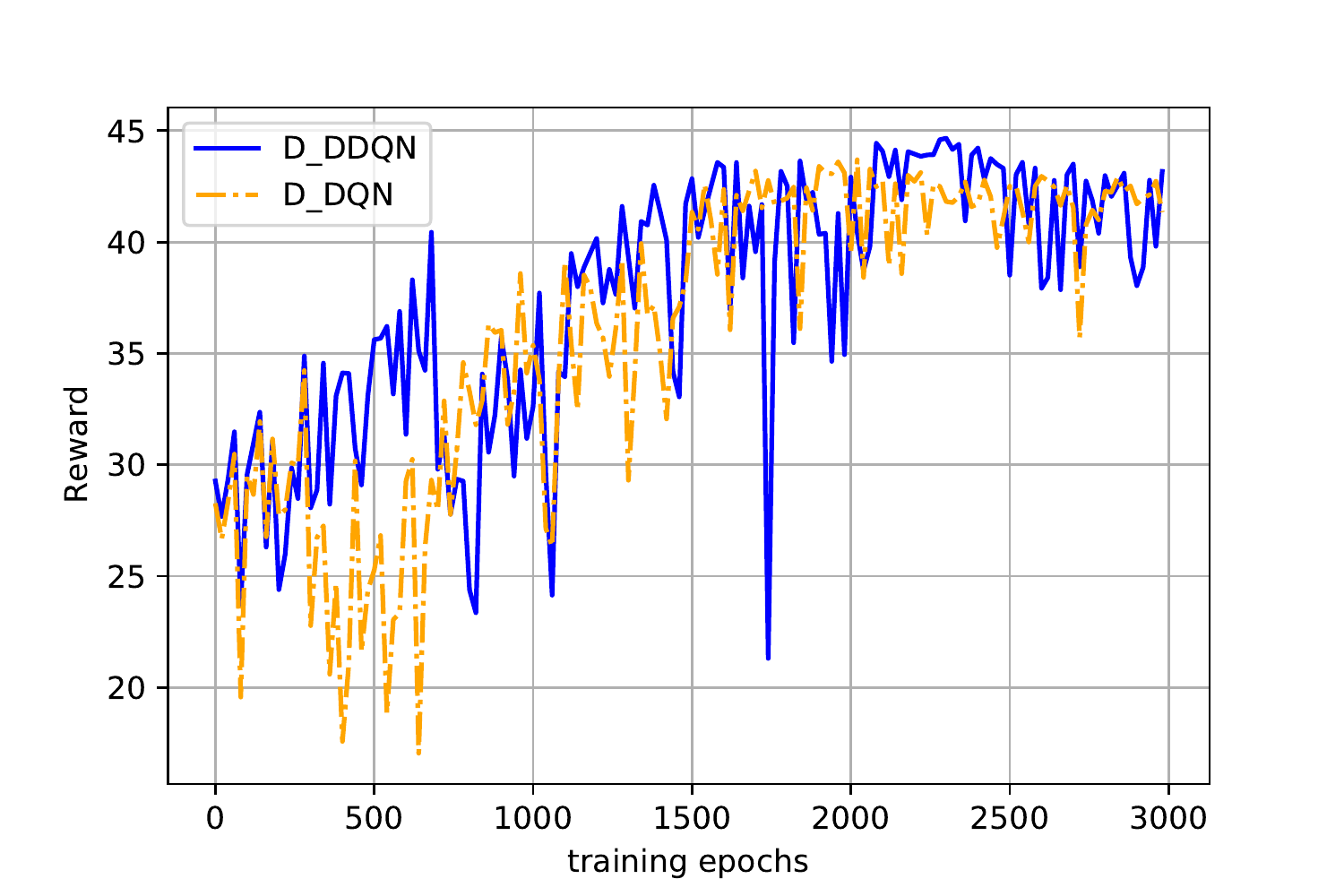}
\centering
\end{minipage}%
}%
\centering
\caption{{Training performance comparison.}}
\label{training_performance}
\end{figure}

Fig. \ref{training_performance}a and Fig. \ref{training_performance}b demonstrate the performance of the proposed DDQN training algorithm by comparing it with the DQN algorithm. It is noted that these two figures shows the loss and the reward of the NN at the BS, and the resource allocation process is performed by MARL. Fig. \ref{training_performance}a illustrates the loss comparison of the DDQN at the BS. The loss of the DDQN algorithm is lower than the DQN training method during the whole training episodes and can achieve faster convergence since it avoids the overestimation problem caused by the DQN based approach. A lower loss of the proposed algorithm leads to a higher training reward since the estimated maximum value of NN is closer to the practical maximum value. As shown in Fig. \ref{training_performance}b, both average rewards per episode of DDQN and DQN algorithm improve as training continues, while the proposed DDQN approach outperforms the DQN based method. Due to the high training loss at the beginning, the DQN based method achieve lower reward than the DDQN based algorithm.

\subsection{Effectiveness and robustness testing}

\begin{figure}[t]
\centering
\includegraphics[width=0.9\columnwidth]{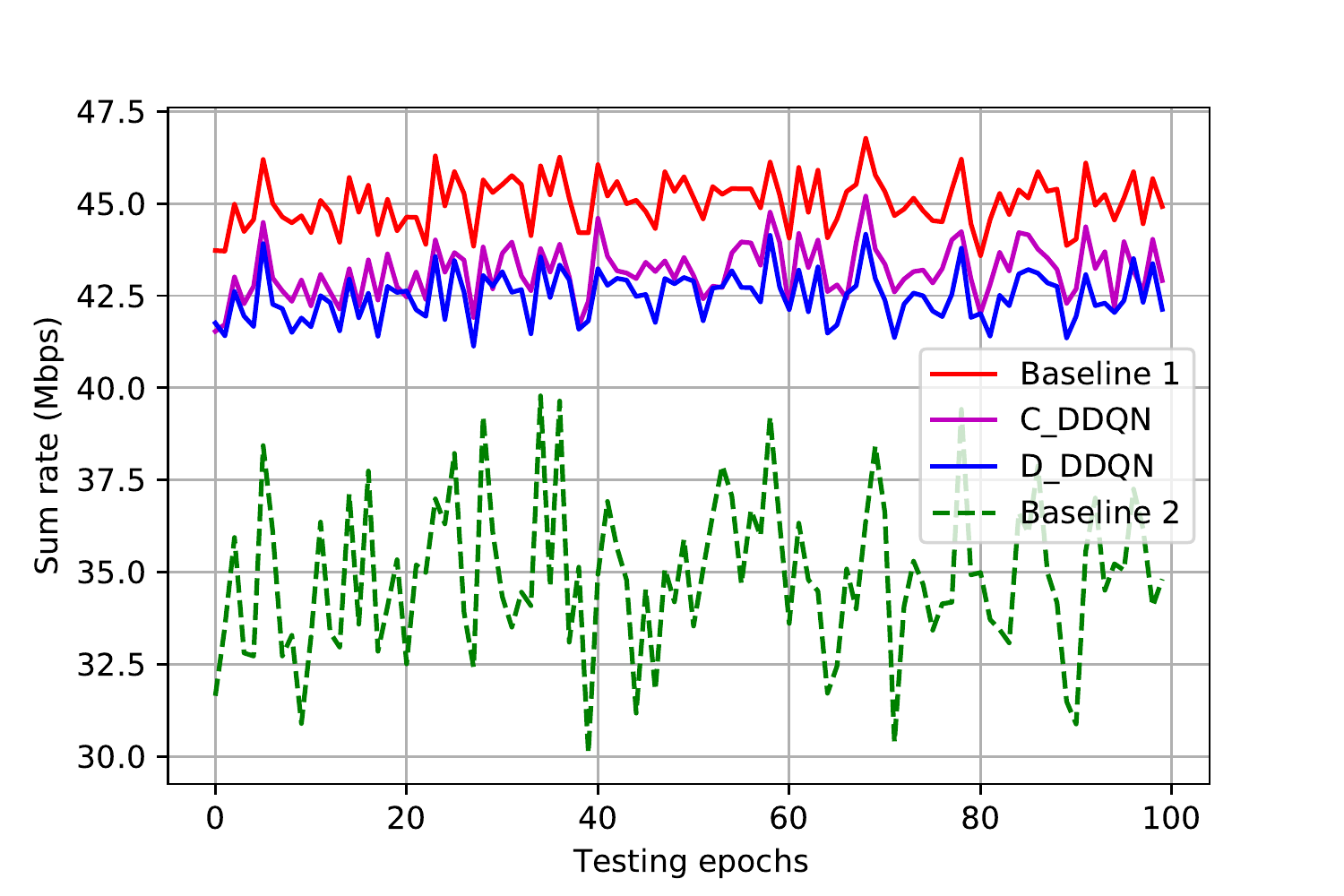}
\caption{Testing performance of proposed algorithm.}
\label{fig5}
\end{figure}

In the testing phase, we verify the effectiveness of the proposed D\_DDQN algorithm. The number of testing step is set as 100 episodes and the exploration rate $\epsilon$ is set as 0 which means the BS always takes the action that has the highest Q-value. As illustrated in Fig. \ref{fig5}, both algorithms can achieve near-optimal performance, reaching about 95\% and 94\% of the optimal solution. Note that the MARL achieves less sum rate than the single-agent RL. This is because it only relies on the local information and could reduce the complexity at the BS.

\begin{figure}[t]
    \centering
    \includegraphics[width=0.9\columnwidth]{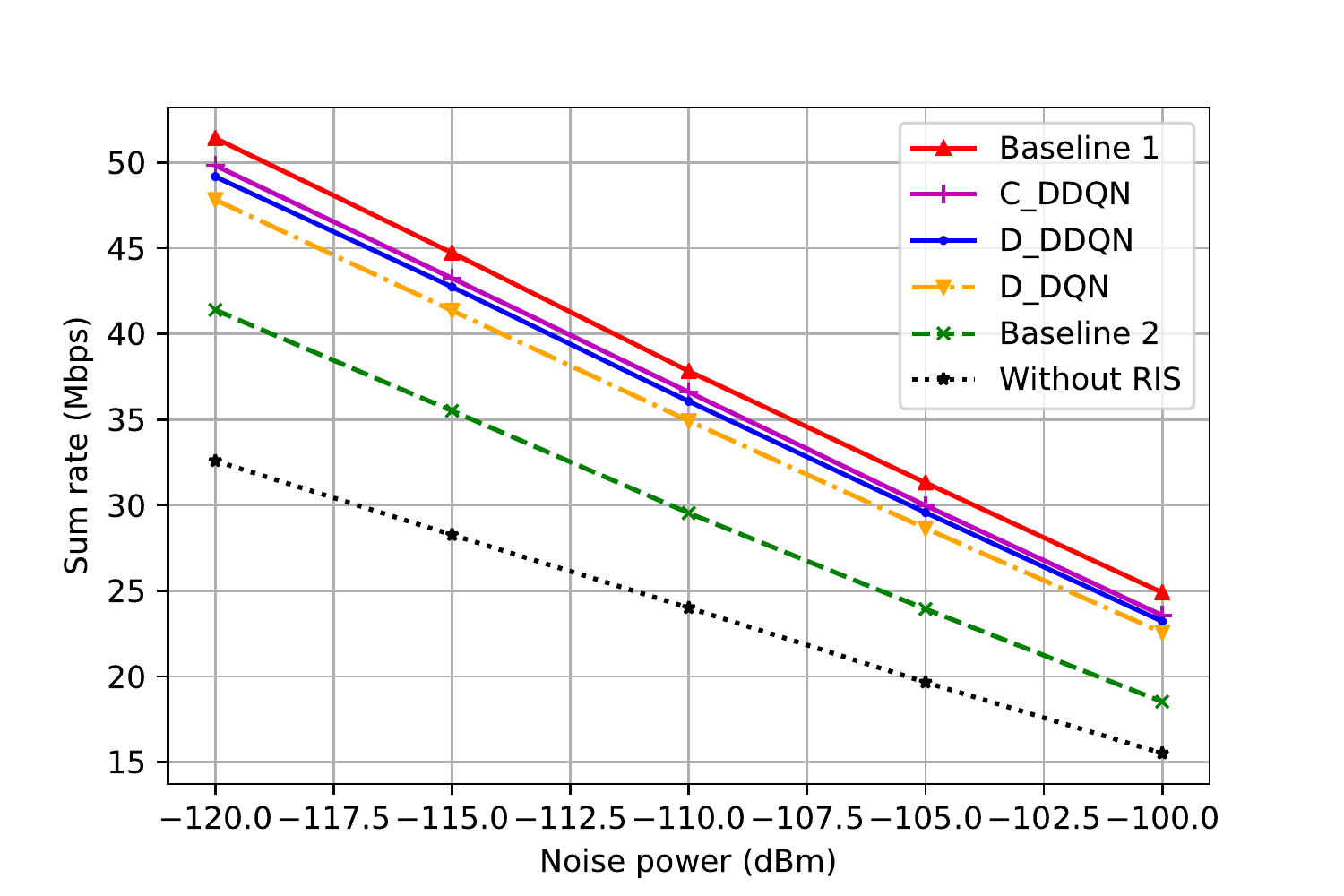}
    \caption{Sum rate comparison over different noise power.}
    \label{fig6}
\end{figure}

We also test the DDQN based algorithm with other benchmarks in various wireless communication scenarios to verify the robustness of the proposed algorithm. Fig. \ref{fig6} shows the sum rate performance under various SINR scenarios. It is observed that the sum rates of all schemes monotonically decrease with higher noise power. As expected, C\_DDQN algorithm and the proposed D\_DDQN algorithm shows near-optimal performance. On the other hand, the random scheme for baseline 2 and transitional wireless communication scheme without the enhancement of RIS behave worse than all of the machine learning optimization methods, providing around only 60\% of the optimal sum rate. Numerical results confirm the performance gain of the proposed algorithm over non RIS schemes, reaching around 50\% improvement of the sum rate performance.

\begin{figure}[t]
\centering
\includegraphics[width=0.9\columnwidth]{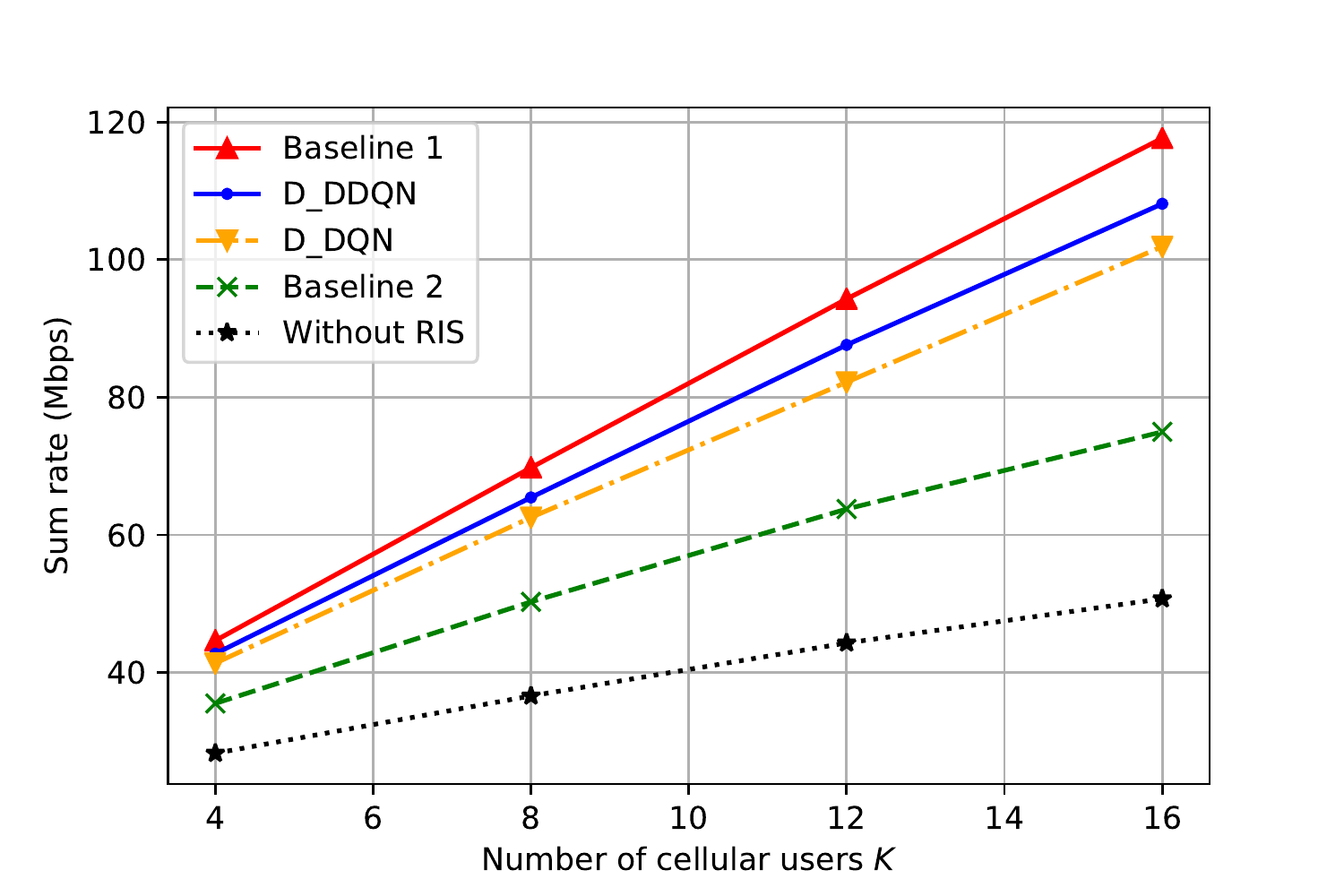}
\centering
\caption{Sum rate over the number of cellular users $K$.}
\label{fig7}
\end{figure}

Fig. \ref{fig7} further demonstrates the sum rate with different number of cellular users. With the increase of accessible channels, the sum rates of all schemes increases. We can observe that the schemes with RIS always outperform without RIS scheme in terms of achieved sum rate, and the gap between the proposed algorithm and the random baseline 2 increases with the number of RBs, which demonstrates the effectiveness of the implementation of RIS. It is also noted that our proposed D\_DDQN algorithm outperforms the D\_DQN and random scheme in all the cases.
\subsection{Impact of the number of the RIS elements}

\begin{figure}[t]
\centering
\includegraphics[width=0.9\columnwidth]{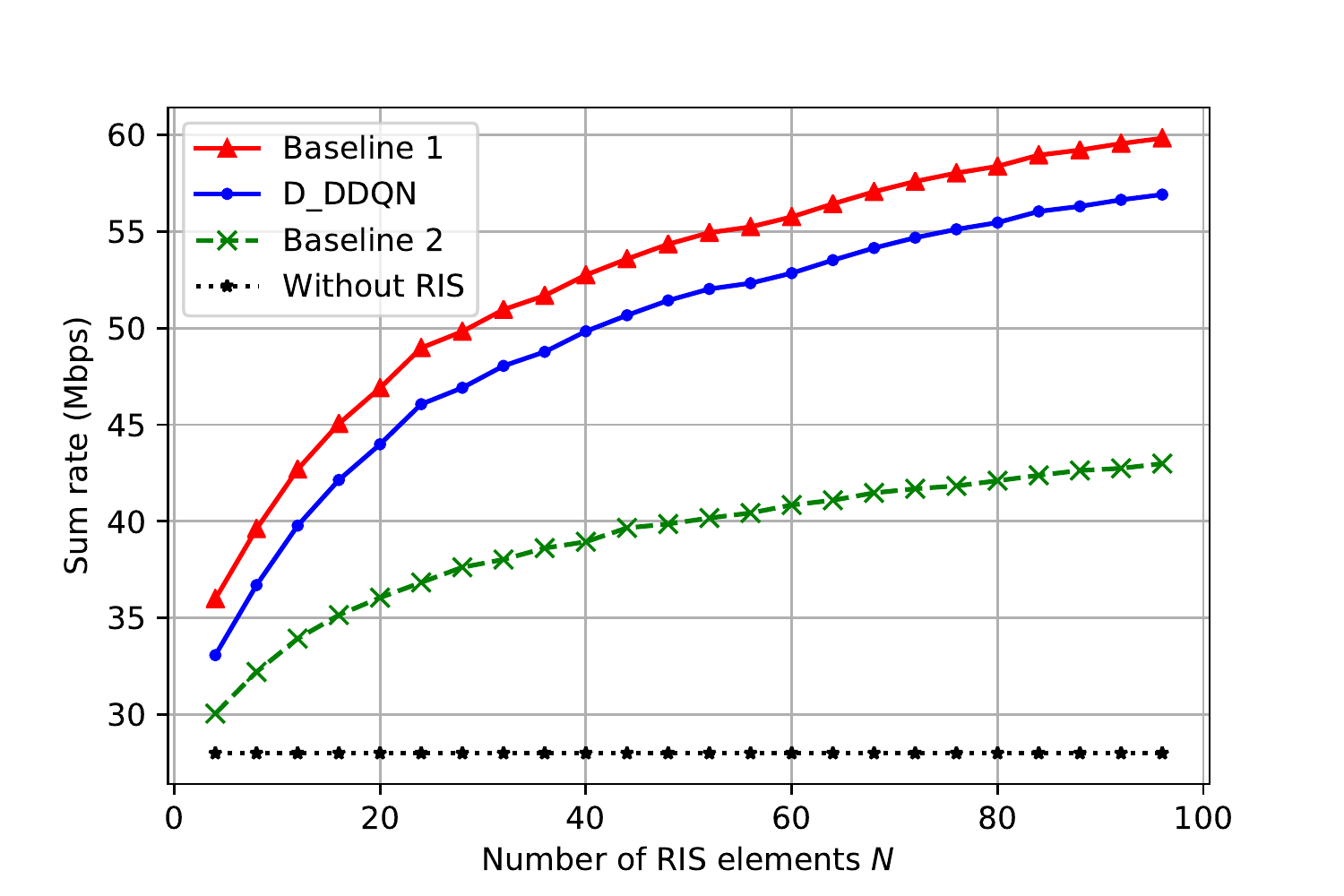} 
\centering
\caption{Sum rate over number of the RIS elements.}
\label{fig8}
\end{figure}

Fig. \ref{fig8} demonstrates the influence of the number of RIS elements on the rate performance. It is observed that the sum rate of the considered network increase with a higher number of RIS elements implemented, while the sum rate for no RIS scheme remains unchanged. Compared to the Baseline 1 with extremely high complexity, the proposed design can achieves 93\% and 95\% of the sum rate performance of the upperbound algorithm when N is 16 and 96, respectively. The suboptimal performance of the proposed algorithm verifies that the undesired channel interference can be effectively suppressed by adequate RIS implement. However, the performance improvement slows down as the number of elements becomes larger, which is caused by the servere interference between cellular networks and D2D communications.

\section{Conclusion}

In this paper, we consider the joint optimization of resource allocation and RIS implementation to maximize the sum rate of the D2D network. To solve the non-convex problem, a novel MARL structure is proposed to perform the channel assignment and RIS optimization. We decouple the joint optimization into sub-problems and reduce the computational pressure at the central BS by decentralized resource allocation. Leveraging the DDQN algorithm, the RIS optimization is performed at the BS centrally. The simulation results demonstrate that the proposed algorithm can achieve near-optimal performance and reduce the computational pressure at the BS significantly. The proposed algorithm outperforms other baseline algorithms under different wireless communication scenarios, verifying the effectiveness and robustness of the DDQN scheme.

\bibliography{Reference}

\end{document}